\RequirePackage{lineno}
\documentclass[aps,twocolumn,showpacs,preprintnumbers,amsmath]{revtex4}
\usepackage{graphicx}
\usepackage{dcolumn}
\usepackage{epstopdf}
\usepackage{color}
\usepackage{bm}
\definecolor{dgreen}{cmyk}{1.,0.,1.,0.2}        
\definecolor{orange}{cmyk}{0.,0.353,1.,0.}    

\def\bea {\begin{eqnarray}}
\def\eea {\end{eqnarray}}

\def\be {\begin{equation}}
\def\ee {\end{equation}}

\begin{document}

\title{Transport Coefficient to Trace Anomaly in the Clustering of Color Sources Approach}
\author{J. Dias de Deus{$^1$}, A. S. Hirsch{$^2$}, C. Pajares{$^3$}, R. P. Scharenberg{$^2$ } and B. K. Srivastava {$^2$}}
\medskip
\affiliation{$^1$CENTRA, Instituto Superior Tecnico, 1049-001 Lisboa, Portugal\\
$^2$Department of Physics and Astronomy, Purdue University, West Lafayette, IN-47907, USA\\ 
$^3$Departamento de Fisica de Particulas, Universidale de Santiago de Compostela and Instituto Galego de Fisica de Atlas Enerxias(IGFAE), 15782 Santiago, de Compostela, Spain}
\bigskip

\date{\today}
\begin{abstract}
 From our previously obtained  shear viscosity to entropy density ratio ($\eta/s$) in the framework of clustering of color sources (Color String Percolation Model: CSPM), we calculate the jet quenching parameter $\hat {q}$ and  trace anomaly $\Delta = (\varepsilon -3\it p)/T^{4}$ as a function of temperature.
It is shown that the scaled $\hat {q}/T^{3}$ is in agreement with the recent JET Collaboration estimates. The inverse of $\eta/s$ is found to  represent $\Delta$. The results for $\Delta$ are in excellent agreement with Lattice Quantum Chromo Dynamics (LQCD) simulations. From the  trace anomaly and energy density $\epsilon$, the equation of state is obtained as a function of temperature and compared with LQCD simulations. It is possible that there is a direct connection between the $\eta/s$ and $\Delta$. Thus the estimate of transport coefficient  $\eta/s$ provides $\hat {q}$ and $\Delta$ as a function of temperature. Both $\Delta$ and $\eta/s$ describe the transition from a strongly coupled QGP to a weakly coupled QGP.

\end{abstract}
\maketitle


\section{Introduction}
The main aim of the heavy ion collision experiments at RHIC and LHC have been to understand the QCD phase transition and properties of the created dense quark matter. Measurements at RHIC and LHC have shown that the matter produced is a strongly coupled QGP and behaves like an almost perfect liquid \cite{brahms,phobos,star,phenix,lhc}. 

Another observation made in central collisions of heavy ions is the suppression of single inclusive hadron spectra at large transverse momentum \cite{jet1}. This phenomena of jet quenching indicates that the produced matter is opaque. Therefore, jet quenching can tell us the properties of the created hot dense matter by the energetic partons passing through the medium and is defined by the jet quenching parameter $\hat{q}$, which is related to another transport coefficient, the shear viscosity to entropy density ratio $\eta/s$ \cite{jet3,jet4}. Both $\eta/s$ and $\Delta$ have been obtained in quasi-particle models \cite{qp1,qp2,qp3,qp4} and the Nambu-Jona-Lasinio model(NJL) as well. 
 
 The phase transition and  $\hat{q}$ have  been studied in the framework of dynamical holographic QCD model \cite{liao}. It is found that both $\hat{q}/T^{3}$ and $\Delta$ peak around the critical temperature. This indicates that $\hat{q}$ can characterize the phase transition \cite{liao}.
  
  In this Letter, we have used our previous work on  $\eta/s$  to evaluate $\hat{q}$ \cite{eos2}. We show that the inverse of $\eta/s$ is a good representation of the trace anomaly $\Delta = (\varepsilon -3\it {p})/T^{4}$.  

\section{ Shear viscosity to entropy density ratio $\eta/s$  }
In our earlier work the shear viscosity to entropy density ratio $\eta/s$ was obtained in the framework of kinetic theory and the string percolation \cite{eos2}. In the Color String Percolation Model (CSPM) the relevant parameter is the transverse string density $\xi= N_{s}S_{1}/S_{n}$ where $N_{s}$ is the number of strings, $S_{1}$ the transverse area of a single string,  $S_{1}= \pi r_{0}^{2}$ and $S_{n}$ the overlap area of the collision, which depends on the impact parameter \cite{pajares1,pajares2}. The following expression was  obtained for $\eta/s$ \cite{eos2}. 
\begin{equation}
\frac {\eta}{s} ={\frac {TL}{5(1-e^{-\xi})}} 
\label{vis}
\end{equation}
where T is the temperature and L is the longitudinal extension of the source $\sim$ 1 $\it {fm} $ \cite{pajares3}.
The temperature is expressed as \cite{pajares3,eos}  
\begin{equation}
T(\xi) =  {\sqrt {\frac {\langle p_{t}^{2}\rangle_{1}}{ 2 F(\xi)}}}
\label{temp}
\end{equation} 
$\langle p_{t}^{2}\rangle_{1}$ is the average transverse momentum squared of particles produced from a single string.
 $F(\xi)$ is the color suppression factor and is related to the percolation density parameter $\xi$ \cite{eos}.
\begin{equation}
F(\xi) = \sqrt {\frac {1-e^{-\xi}}{\xi}}
\label{fxi}
\end{equation} 
The connection between $\xi$ and the temperature $T(\xi)$ involves the Schwinger mechanism (SM) for particle production. 
The Schwinger distribution for massless particles is expressed in terms of $p_{t}^{2}$ \cite{swinger,wong}
\begin{equation}
dn/d{p_{t}^{2}} \sim e^{-\pi p_{t}^{2}/x^{2}}
\end{equation}
where the average value of the string tension is  $\langle x^{2} \rangle$. The tension of the macroscopic cluster fluctuates around its mean value because the chromo-electric field is not constant.
The origin of the string fluctuation is related to the stochastic picture of 
the QCD vacuum. Since the average value of the color field strength must 
vanish, it can not be constant but changes randomly from point to point \cite{bialas}. Such fluctuations lead to a Gaussian distribution of the string tension, which transforms SM into the thermal distribution \cite{bialas}
\begin{equation}
dn/d{p_{t}^{2}} \sim e^{(-p_{t} \sqrt {\frac {2\pi}{\langle x^{2} \rangle}} )}
\end{equation}
with $\langle x^{2} \rangle$ = $\pi \langle p_{t}^{2} \rangle_{1}/F(\xi)$. 

The string percolation density parameter $\xi$ which characterizes the percolation clusters also determines the temperature of the system. In this way at $\xi_{c}$ = 1.2 the connectivity percolation transition at $T(\xi_{c})$ models the thermal deconfinement transition.
We adopt the point of view that the experimentally determined universal chemical freeze-out temperature ($\it T_{f}$) is a good measure of the phase transition temperature, $T_{c}$ \cite{braunmun}. ${\langle p_{t}^{2}\rangle_{1}}$ is calculated at $\xi_{c}$ = 1.2 using the $\it {T_{f}}$ = 167.7 $\pm$ 2.6 MeV \cite{bec1}. This gives $ \sqrt {\langle {p_{t}^{2}} \rangle _{1}}$  =  207.2 $\pm$ 3.3 MeV which is close to  $\simeq$ 200 MeV used in a previous calculation of the percolation transition temperature \cite{pajares3}.

\begin{figure}[thbp]
\centering        
\vspace*{-0.2cm}
\includegraphics[width=0.55\textwidth,height=3.0in]{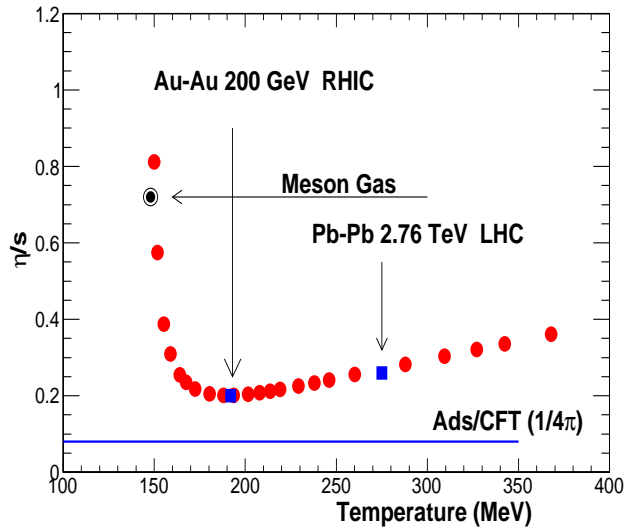}
\vspace*{-0.5cm}
\caption{(Color online) $\eta/s$ as a function of temperature T using Eq. (\ref{vis}) \cite{eos2}. Au+Au at 200 GeV for 0-10$\%$ centrality is shown as solid blue square \cite{eos2}. The estimated value for Pb+Pb at 2.76 TeV for 0-5$\%$ centrality  is shown as a solid blue square.  The meson gas value for $\eta/s$ $\sim$ 0.7 is shown as solid black circle at T $\sim$ 150 MeV \cite{prakash}. The lower bound shown is given by the AdS/CFT \cite{kss}.} 
\label{visfig}
\end{figure}

Figure~\ref{visfig} shows $\eta/s$ as a function of the temperature \cite{eos2}. The lower bound shown in Fig.~\ref{visfig} is given by the AdS/CFT conjecture \cite{kss}. The results from Au+Au at 200 GeV and Pb+Pb at 2.76 TeV collisions show that the $\eta/s$ value is 2.5 and 3.3 times the KSS bound \cite{kss}. The CSPM values of $\eta/s$ as obtained using Eq.~(\ref{vis}) can be lowered by 15 $\%$ due to the slightly lower value of $L$ at higher densities.

 In CSPM  $\eta/s$ is not needed to reproduce the elliptic flow and higher harmonics. The cluster formed by the strings has generally an asymmetric form in the transverse plane and acquires dimensions comparable to the nuclear overlap. This azimuthal asymmetry is at the origin of the elliptic flow in CSPM. The partons emitted at some point inside the cluster have to pass through the strong color field before appearing on the surface. The results of the simulation for different harmonics are in reasonable agreement with the experimental data on the $p_t$ and centrality dependencies ~\cite{flow1,flow2}.

It has been observed that $\eta/s$ has  minimum in the phase transition region in systems like helium, nitrogen, water, and many other substances \cite{larry,koch}. 
Thus it shows the location of the transition from hadrons to quarks and gluons or crossover in QCD. Our results show that the fall and rise of $\eta/s$ as a function of temperature is necessary to keep $v_{2}(p_{t})$ almost constant in going  from RHIC to LHC energies as observed in ref.~ \cite{plumari}.

\section{ $\eta/s$ to  Scaled jet quenching parameter $\hat{q}/T^{3}$}
The small shear viscosity of the QGP implies strong jet quenching. It has been suggested that the $\hat{q}$ can also be used to measure the coupling strength of the medium. The shear viscosity $\eta$ of a weakly coupled plasma can be related to the transport parameter for a thermal parton $\hat{q}$ \cite{jet3,jet4}. 
\begin{equation}
\frac {\eta}{s} \approx {\frac{3}{2} \frac {T^{3}}{\hat{q}}}
\end{equation}
 The relation associates a small ratio of $\eta/s$ to a large value of  $\hat{q}$. A large amount of theoretical work has been done to extract the jet transport parameter from jet quenching at RHIC and LHC energies \cite{jet1,jet3,jet4,jetqm}. The latest study by the JET Collaboration has extracted or calculated $\hat{q}$ from five different approaches to the parton energy loss in a dense medium.
 The evolution of bulk medium in the study was given by 2+1D or 3+1D 
hydrodynamic models with the initial temperatures of $T_{RHIC}^{Hydro}$ = 346-373 MeV  and $T_{LHC}^{Hydro}$ = 447-486 MeV for most central Au+Au collisions at 
$\sqrt{s_{NN}}$ = 200 GeV and Pb+Pb collisions at $\sqrt{s_{NN}}$ = 2.76 TeV respectively. 
 The variation of $\hat{q}$ values between different models can be considered as theoretical uncertainties. One therefore can extract its range of values at RHIC and LHC \cite{jet1,jetqm}.
\begin{equation}
\frac{\hat{q}}{T^{3}} \approx \{^{4.5\pm 1.3 \   at \ RHIC}_ {3.7\pm1.4 \    at \ LHC},
\label{qeq}
\end{equation}
at the highest temperatures reached in the most central Au+Au collisions at RHIC and Pb+Pb collisions at LHC. The corresponding absolute values for 
$\hat{q}$($GeV^{2}/fm$) for a 10 GeV quark jet are,  
\begin{equation}
\hat{q} \approx \{^{1.2\pm 0.3 \   }_ {1.9\pm0.7 } {^{T= 370 MeV}_{ T=470 MeV}},
\label{qeq2}
\end{equation}
at an initial time $\tau_{0} = 0.6 fm/c$. 
\begin{figure}[thbp]
\centering        
\vspace*{-0.2cm}
\includegraphics[width=0.55\textwidth,height=3.0in]{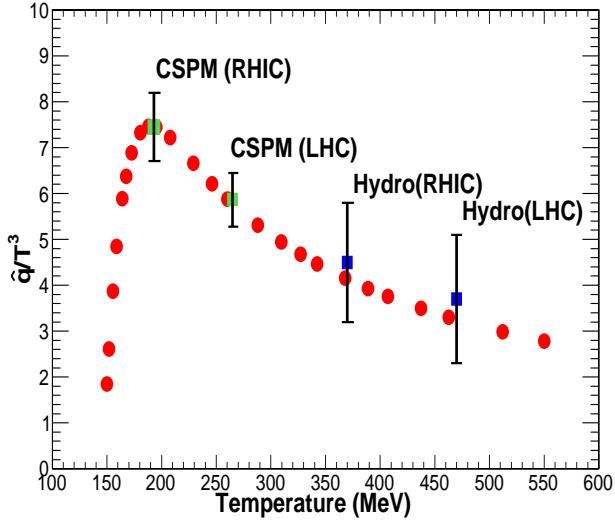}
\vspace*{-0.5cm}
\caption{ (Color online) Scaled jet quenching parameter $\frac{\hat{q}}{T^{3}}$ as a function of the temperature. The values shown in solid blue squares Hydro(RHIC) and Hydro(LHC) are given by Eq. (\ref{qeq}). The CSPM values are shown in green solid squares (CSPM(RHIC) and CSPM(LHC) for temperatures $\sim$193 and $\sim$262 MeV at RHIC and LHC energies, respectively \cite{eos2}.}
\label{qhat}
\end{figure} 
The temperature dependence of scaled jet transport parameter $\frac {\hat{q}}{T^{3}}$ is shown in Fig. \ref{qhat}. The CSPM values are shown as solid green squares while the theoretical values are shown as blue squares. It is observed that CSPM values are in agreement with the JET Collaboration results. 
\section{ $\eta/s$ and Trace anomaly $\Delta$}
\begin{figure}[thbp]
\centering        
\vspace*{-0.2cm}
\includegraphics[width=0.55\textwidth,height=3.0in]{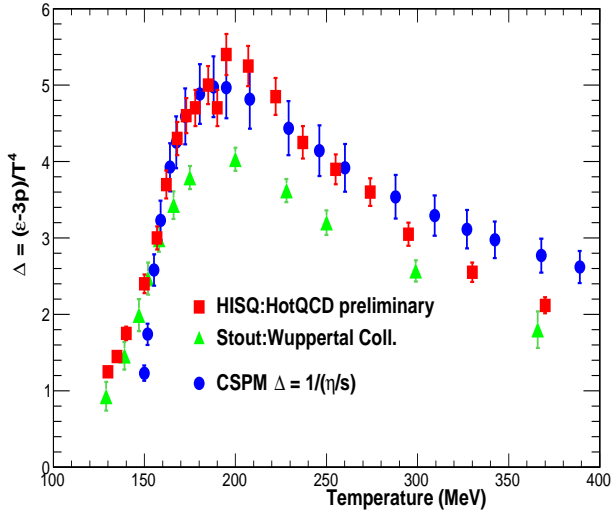}
\vspace*{-0.5cm}
\caption{(Color online) The trace anomaly $\Delta =(\varepsilon-3p)/T^{4}$ vs temperature. Red solid squares are from HotQCD Collaboration \cite{lattice12}. Green triangles are from Wuppertal Collaboration \cite{wuppe}. The CSPM results are shown as solid blue circles and is obtained as $\Delta = 1/(\eta/s)$ \cite{IS2013}.} 
\label{trace}
\end{figure}

The trace anomaly ($\Delta$) is the expectation value of the trace of the energy-momentum tensor, $\langle \Theta_{\mu}^{\mu}\rangle = (\varepsilon-3p)$, which measures the deviation from conformal behavior and thus identifies the interaction still present in the medium \cite{cheng}. We consider the $\it ansatz$ that inverse of $\eta/s$ is equal to trace anomaly $\Delta$. $\eta/s$ is in quantitative agreement with $(\varepsilon-3p)/T^{4}$ over a wide range of temperatures \cite{cpod13,IS2013} 
 This result is shown in Fig.~\ref{trace}. The minimum in $\eta/s \sim 0.20$ determines the peak of the interaction measure $\sim$ 5 in agreement with the recent HotQCD values \cite{lattice12}. This happens  at the critical temperature of $T_{c} \sim 175$ MeV. Figure~\ref{trace} also shows the results from Wuppertal Collaboration \cite{wuppe}. As mentioned earlier in Sec.II that $\eta/s$ can be lower by $\sim$ 15$\%$ at high densities. This has the effect of increasing $\Delta$. The peak value of $\Delta$ = 4.88 increases to 5.61. 

The maximum in $\Delta$ corresponds to the minimum in $\eta/s$. Both $\Delta$ and $\eta/s$ describe the transition from a strongly coupled QGP to a weakly coupled QGP. 

We are not aware of any theoretical work which directly relates the trace anomaly with the shear viscosity to entropy density ratio. However, the bulk viscosity $\zeta$ is related to both $\Delta$ and $\eta$ \cite{teany}. A detailed study based on low energy theorms and the lattice result for $\Delta$ shows that $\zeta/s$ rises very fast close to the critical temperature in such a way that its value at temperatures higher than $T >1.1  T_{c}$ is quite negligible \cite{karsch}. 

It was observed that $\zeta$ scales as $\alpha^{4}_{s} \eta$ where $\alpha_{s}$ is the coupling constant. The trace anomaly $\Delta$ is proportional to $\alpha^{2}_{s}$ \cite{teany}. There are many other works in which $\Delta$ and $\eta/s$ have been obtained separately \cite{qp2,plumari2,bluhm2,marty}.
%
\section{Equation of State EOS : The sound velocity $C_{s}^{2}$}
\begin{figure}[thbp]
\centering        
\vspace*{-0.2cm}
\includegraphics[width=0.55\textwidth,height=3.0in]{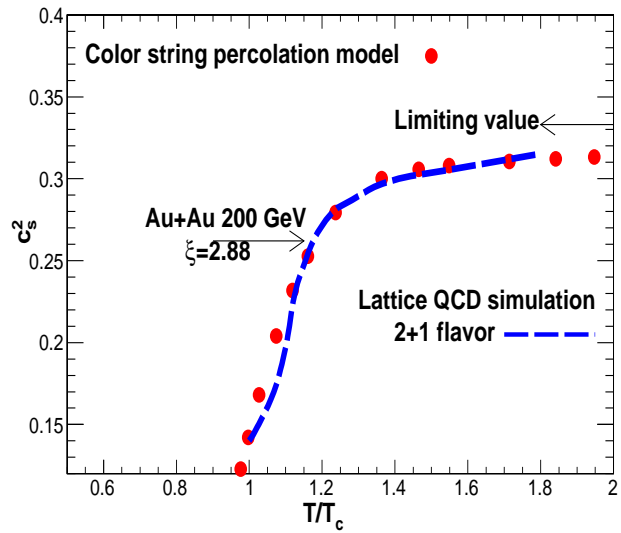}
\vspace*{-0.5cm}
\caption{(Color online) The speed of sound from CSPM (red circles) and Lattice QCD-p4 versus  $T/T_{c}$ (blue dash line) \cite{hotqcd}.} 
\label{cs2}
\end{figure}

An analytic expression for the equation of state, the sound velocity $C_{s}^{2}$  is obtained in CSPM.
After the initial temperature $ T > T_{c}$ the CSPM perfect fluid may expand according to Bjorken boost invariant 1D hydrodynamics \cite{bjorken}. The input parameters the initial temperature T, the initial energy density $\varepsilon$, and the trace anomaly $\Delta$ are determined by data. The Bjorken 1D expansion gives the sound velocity   
\begin{eqnarray}
\frac {1}{T} \frac {dT}{d\tau} = - C_{s}^{2}/\tau  \\
\frac {d\varepsilon}{d\tau} = -T s/\tau 
\end{eqnarray}
where $\varepsilon$ is the energy density, s the entropy density, $\tau$ the proper time, and $C_{s}$ the sound velocity. Since $s= \varepsilon + p/T$ and $p = (\varepsilon-\Delta T^{4})/3$ one gets
\begin{equation}
\frac {dT}{d\varepsilon} s = C_{s}^{2} 
\end{equation}
From above equations $C_{s}^{2}$ can be expressed in terms of $\xi$

 \[C_{s}^{2} = (-0.33)\left(\frac {\xi e^{-\xi}}{1- e^{-\xi}}-1\right) +\] 
 \begin{equation}  (\Delta/3)\left(\frac {0.019}{1-e^{-\xi}}\right)
   \left(\frac {\xi e^{-\xi}}{1- e^{-\xi}}-1 \right)
\label{sound}
\end{equation}
Since there is no direct way to obtain pressure in the CSPM, we have made the assumption that $\Delta = (\varepsilon- 3 P) \approx 1/(\eta/s)$ . Fig.~\ref{cs2} shows a plot of
 $C_{s}^{2}$ as a function of $T/T_{c}$. It is observed that the CSPM results are in very good agreement with the lattice calculations \cite{hotqcd}. This suggests that the $\Delta$ can be approximated to  $1/(\eta/s)$.

\section{Discussion}
We have shown that the inverse of the shear viscosity to entropy density ratio is able to give good description of the trace anomaly. The jet quenching transport coefficient $\hat{q}$ is also obtained using the relation with $\eta/s$. It is observed that  scaled jet quenching parameter shows a pronounced maximum close to the critical temperature as seen in the trace anomaly. This indicates that jet quenching parameter can characterize the phase transition \cite{liao,shuryak}. It has been also shown, with in a phenomenological quasi-particle approach, that  trace anomaly has a pronounced peak near the critical temperature \cite{bluhm,casto}.

 The clustering of color sources has shown that the determination of $\eta/s$ as a function of temperature is an important quantity that relates to another transport coefficient, $\hat{q}$ and the trace anomaly $\Delta$. The main assumption of the present approach is  that the inverse of $\eta/s$ represents the trace anomaly,  $\Delta = (\varepsilon -3{\it p)/T^{4}}$. The clustering of color sources (percolation) provides us with a microscopic partonic picture that connects the transport properties of the QGP to its thermodynamics.

\section{Acknowledgment}
This research was supported by the Office of Nuclear Physics within the U.S. Department of Energy  Office of Science under Grant No. DE-FG02-88ER40412. 
%

\end{document}